\documentclass[prd,a4paper,tightenlines,eqsecnum,nofootinbib,preprint,floatfix,showpacs]{revtex4}
\usepackage{epsfig}
\usepackage{graphicx}
\usepackage{bm}

\def\Tr{\text{Tr}\,}

\def\Eq#1{Eq.~(\ref{#1})}

\def\<{\langle}
\def\>{\rangle}

\begin{document}

\mbox{} \hfill OUTP-05-06P~~\\

\vspace*{1.0in}

\title{The pressure of the $SU(N)$ lattice gauge theory at large--$N$ \\
\vspace*{0.25in}}
\author{Barak Bringoltz and Michael Teper\\
\vspace*{0.25in}}

\affiliation{Rudolf Peierls Centre for Theoretical physics,
University  of  Oxford,\\ 
1 Keble Road, Oxford, OX1 3NP, UK \\
\vspace*{0.6in}}

\begin{abstract}
We calculate bulk thermodynamic properties, such as the pressure,
energy density, and entropy, in SU(4) and SU(8) lattice gauge
theories, for the range of temperatures $T \leq 2.0T_c$ and 
$T\leq 1.6T_c$ respectively. We find that the $N=4,8$ 
results are very close to each other, and to what one finds
in SU(3), and are far from the asymptotic free-gas value. We conclude 
that any explanation of the high-$T$ pressure (or entropy) deficit
must be such as to survive the $N\to\infty$ limit. We give some
examples of this constraint in action and comment on what this
implies for the relevance of gravity duals.

\end{abstract}

\maketitle

\section{Introduction}
\label{sec:intro}

The thermodynamic properties of Quantum Chromodynamics (QCD), besides
being of fundamental interest, are currently at the centre of 
intense experimental research. One of the most interesting phenomena 
has to do with the range of temperatures, $T$, above the
phase transition (or crossover) at $T=T_c$, where the theory
deconfines and chiral symmetry is restored. Traditionally, the 
description of this transition assumed that the hadronic phase gives 
way to a plasma, whose physical degrees of freedom are weakly
interacting quarks and gluons. Recent experimental results have,
however, challenged this `simple'  picture (for example see 
\cite{Heinz:2004ar} 
and references therein), and point to a picture of the `plasma' 
as a very good fluid in the accessible range of $T$ above $T_c$. 
In fact, numerical lattice results had already demonstrated the 
inadequacy of the simple quark-gluon plasma picture some time ago.
Such lattice calculations, both for the pure gauge case 
\cite{Boyd:1996bx}
and with different kinds of fermions 
\cite{Engels:1996ag},
found a large deficit in the pressure and entropy as compared 
to the Stephan-Boltzmann predictions for a free gluon gas (for 
pure glue), which remained at the level of more than $10\%$ 
even at temperatures as high as $T\sim 4T_c$. 
Further evidence that points in the same direction is the survival 
of hadronic states above $T_c$, as seen in recent lattice 
simulations (for example see 
\cite{Petreczky:2004xs} 
and references therein).

These lattice calculations, and more recent experimental 
observations, have attracted considerable attention
(see e.g. 
\cite{Karsch:2001cy}
for a review).
Approaches have ranged from modeling the system in terms of 
noninteracting quasi-particles with the quantum numbers of 
quarks and gluons but with temperature dependent masses 
\cite{Levai:1997yx,Peshier:1995ty},
to using higher order perturbation theory (restricted by infrared 
divergences), sometimes including  nonperturbative contributions 
on the dimensionally reduced $3D$ Euclidean lattice
\cite{Schroder:2004gt}, large re-summations (e.g. \cite{Blaizot:2003tw} and references therein), or, more recently, a description
\cite{Shuryak:2004tx} in terms of a large number of loosely bound states that survive
deconfinement and come in various representations of the 
gauge and flavor groups, and where one can use for example
the lattice masses measured in \cite{Petreczky:2001yp}.

In this paper we ask whether this pressure (and entropy) deficit
is a dynamical feature not just of SU(3) but of all SU($N$)
gauge theories -- and in particular whether it survives the
$N\to \infty$ limit. In this limit the theory becomes considerably
simpler, although not (yet) analytically soluble, and so what happens 
there should strongly constrain the possible dynamics 
underlying the phenomenon. For example, in that limit supersymmetric SU($N$)
gauge theories become dual to weakly coupled gravity models,
and in that context we recall the frequently mentioned prediction 
\cite{Gubser:1998nz},
that the pressure in the strong-coupling limit of the ${\cal N}= 4$ 
and $N=\infty$ supersymmetric gauge theory is $3/4$ of its
Stephan-Boltzmann value, which is similar to the deficit, referred
to above, that one finds in the non-supersymmetric case.

To address this question we calculate the pressure for $T\leq 2T_c$
in $SU(4)$, and $SU(8)$ lattice gauge theories and compare
the results to similar SU(3) calculations available in the literature
(which we supplement where it is useful to do so). Recent calculations 
of various  properties of SU($N$) gauge theories 
\cite{Teper:2004pk}
have demonstrated that SU(8) is in fact very close to SU($\infty$) 
for most purposes and have provided information on
the location, $\beta_c$, of the deconfining transition for various 
$L_t$ and $N$
\cite{Lucini:2002ku,Lucini:2003zr}.
Thus our calculations should provide us with an accurate picture 
of what happens to the pressure at $N=\infty$.

In the next Section we summarise the lattice setup, the
relevant thermodynamics, and provide numerical checks that
our system is large and homogeneous enough for our
thermodynamic relations to be appropriate. We then present 
our results for the pressure, entropy and related quantities.
We discuss the implications of our findings in the concluding section.

\section{Lattice set up and methodology}
\label{sec:lattice}

The theory is defined on a discretised periodic Euclidean four 
dimensional space-time with $L^3_s\times L_t$ sites. Here $L_{s,t}$ 
is the lattice extent in the spatial and Euclidean time 
directions. The partition function 
\begin{equation}
Z(T,V) = \sum_s \exp \left \{-\frac{E_s}{T} \right \} 
= 
\exp\left\{-\frac{F}{T}\right\}
=
\exp\left\{-\frac{fV}{T}\right\}
\label{eq:ZF}
\end{equation}
defines the free energy $F$ and the free energy density, $f$,
and can be expressed as a Euclidean path integral
\begin{equation}
Z(T,V)=\int DU \exp{\left( -\beta S_{\rm W}\right)}.
\label{eq:Z}
\end{equation}
Here $T=(aL_t)^{-1}$ is the temperature and $V=(aL_s)^3$ is the
spatial volume. When we change $\beta$, so as to change 
the lattice spacing $a(\beta)$, we change both $T$ and $V$,
if $L_s$ and $L_t$ are kept fixed.
In the large--$N$ limit, the 't Hooft coupling $\lambda=g^2N$ 
is kept fixed, and so we must scale $\beta=2N^2/\lambda \propto N^2$
in order to keep the lattice spacing fixed in that limit. 
We use the standard Wilson action $S_{\rm W}$ given by
\begin{equation}
S_{\rm W}=\sum_P \left[ 1- \frac1N {\rm Re}\Tr{U_P} \right].
\label{eq:SW}
\end{equation}
Here $P$ is a lattice plaquette index, and $U_P$ is the plaquette 
variable obtained by multiplying link variables along the circumference 
of a fundamental plaquette. We perform Monte-Carlo simulations, 
using the Kennedy-Pendelton heat bath algorithm for the link 
updates, followed by five over-relaxations of all the $SU(2)$ 
subgroups of $SU(N)$.

\subsection{The method used}

In lattice calculations of bulk thermodynamics, one can choose
to use either the ``integral'' method (e.g. \cite{Boyd:1996bx}) 
or the  ``differential''  method (e.g. \cite{Gavai:2004se} or a new variant \cite{Gavai:2005da}) or 
one can attempt a direct evaluation of the density of states 
(e.g. \cite{Bhanot:1986hi}). We choose to use the first of these 
methods since the numerical price involved in using larger values 
of $N$ drives us to smaller $L_t$, which means that the lattice 
spacing is too coarse (about $0.15{\rm fm}$) for the differential 
method. We have performed preliminary checks for the applicability 
of the Wang-Landau algorithm \cite{WangLandau} for the evaluation 
of the density of states in the $SU(8)$ gauge theory, but found it 
numerically too costly for the present work.

The properties we will concentrate on are the pressure $p$, the 
energy density per unit volume $\epsilon$, and the entropy $S$,
as a function of temperature. These are given by
\begin{eqnarray}
p &=& T \frac{\partial}{\partial V} \log Z(T,V)
= \frac{T}{V}\log Z(T,V) = -f , \label{eq:P} \\
\epsilon &=& \frac{T^2}{V} 
\frac{\partial}{\partial T} \log Z(T,V), \label{eq:e} \\
\frac{S}{V} &=& \frac{\epsilon - f}{T}
= \frac{\epsilon + p}{T}. \label{eq:s}
\end{eqnarray}
where the second equality in the first and last lines follows
if the system is large and homogeneous, i.e. if $V$ is large enough. 
In addition it is useful to consider the quantity
\begin{equation}
\Delta \equiv \epsilon-3p
= T^5 \frac{\partial}{\partial T} \frac{p}{T^4}
\label{eq:delta_e}
\end{equation}
which vanishes for an ideal gluon plasma. Again the second
equality requires a large enough $V$. To calculate the
pressure at temperature $T$ in a volume $V$ with lattice
cut-off $a(\beta)$, we express $\log Z$ in the integral form 
\begin{equation}
p(T) = \frac{T}{V}\log Z(T,V)
=
 \frac1{a^4(\beta)L_s^3L_t}
\int^\beta_{\beta_0}  d\beta^\prime 
\frac{\partial\log Z }{\partial \beta^\prime} 
\label{eq:integral}
\end{equation}
(There is in general an integration constant, but it will disappear 
when we regularise the pressure later on in this section.)
This integral form is useful because it is easy to see from 
Eqs.~(\ref{eq:Z},\ref{eq:SW}) that 
\begin{equation}
\frac{\partial\log Z }{\partial \beta}
=
-\langle S_W \rangle
=
N_p \langle u_p \rangle
\label{eq:pZup}
\end{equation}
where $N_p=6L_tL^3_s$ is the total number of plaquettes and
$u_p \equiv {\rm Re}\Tr{U_P}/N$. So the pressure can be
obtained by simply integrating the average plaquette over $\beta$. 
This pressure has been defined relative to that of the
unphysical `empty' vacuum and will therefore be ultraviolet 
divergent in the continuum limit. To remove this divergence we 
need to define the pressure relative to that of a more
physical system. We shall follow convention and subtract 
from $p(T)$ its value at $T=0$, calculated with the same 
value of the cut-off $a(\beta)$.
Thus our pressure will be defined with respect to its
$T=0$ value. Doing so we obtain from Eq.~(\ref{eq:pZup},
\ref{eq:integral})
\begin{equation}
a^4[ p(T) - p(0) ]
=
6 \int^\beta_{\beta_0}  d\beta^\prime 
(\langle u_p \rangle_T - \langle u_p \rangle_0).
\label{eq:pint1}
\end{equation}
where $\langle u_p \rangle_0$ is calculated on some
$L^4$ lattice which is large enough for it to be 
effectively at $T=0$. 
We replace $ p(T) - p(0) \to p(T)$, where from now on it
is understood that $p(T)$ is defined relative to its value
at $T=0$, and we use  $T=(aL_t)^{-1}$ to rewrite 
\Eq{eq:pint1} as
\begin{equation}
\frac{p(T)}{T^4}
=
6 L^4_t \int^\beta_{\beta_0} d\beta^\prime 
(\langle u_p \rangle_T - \langle u_p \rangle_0).
\label{eq:pint2}
\end{equation}
We remark that when our $L_s^3L_t$ lattice is in the confining
phase, then $\langle u_p \rangle$ is essentially independent 
of $L_t$ and takes the same value as on a $L^4_s$ lattice (see 
below). This should
become exact as $N\to\infty$ but is accurate enough even for
SU(3). Thus as long as we choose $\beta_0$ in \Eq{eq:pint2}
such that $a(\beta_0)L_t > 1/T_c$ then the integration constant,
referred to earlier, will cancel.

Finally, we evaluate $\Delta$ in \Eq{eq:delta_e} as follows:
\begin{eqnarray}
\Delta/T^4
&=& T\frac{\partial}{\partial T} \frac{p}{T^4} \\
&=& \frac{\partial\beta}{\partial \log T}
\frac{\partial}{\partial \beta} \frac{p}{T^4} \\ 
&=& 6 L_t^4 
(\langle u_p(\beta) \rangle_0 - \langle u_p(\beta) \rangle_T)
\times
\frac{\partial \beta}{\partial \log (a(\beta))}. 
\label{eq:final_delta}
\end{eqnarray}
To evaluate ${\partial \log (a(\beta))}/{\partial \beta}$ we
can use calculations of the string tension, $\sigma$,
in lattice units. For example, in 
\cite{Lucini:2005vg}
the calculated values of $a\surd\sigma$ are interpolated in
$\beta$ for various $N$ and one can take the derivative of
the interpolated form to use in \Eq{eq:final_delta}.
One could equally well use the calculated mass gap or
the deconfining temperature. All these choices will give
the same result up to modest $O(a^2)$ differences.

\subsection{Average plaquette}

We see from the above that what we need to do is to calculate
average plaquettes closely enough in $\beta$ so as to be able 
to perform the numerical integration in $\beta$. And we need the 
average plaquettes not only on the $L_tL^3_s$ lattice but also on
a reference `$T=0$' $L^4$ lattice at each value of $\beta$. 
However we mostly need values for $\beta \geq \beta_c$, where 
$a(\beta_c)L_t=1/T_c$, since $p(T)-p(0) \simeq 0$ once $T<T_c$. 

We performed calculations in $SU(4)$ on $16^3 5$ lattices and 
in $SU(8)$ on $8^3 5$ lattices for a range of 
$\beta$ values corresponding to $T/T_c \in [0.89,1.98]$ for 
$SU(4)$, and to $T/T_c\in [0.97,1.57]$ for $SU(8)$. Since we use 
$L_t=5$, while the data for $SU(3)$ in \cite{Boyd:1996bx} is 
for $L_t=4,6,8$, we also performed simulations for $SU(3)$ 
on $20^3 5$ lattices with $T/T_c\in [1,2]$.
The results are presented in Tables~\ref{table1}--\ref{table3}.

In addition to the finite $T$ calculations we have performed
`$T=0$' calculations on $20^4$ lattices for SU(3), and on
$16^4$ lattices for SU(4). These have the advantage of being
on the same spatial volumes as the corresponding finite $T$ 
calculations, and we know from previous calculations 
\cite{Lucini:2001ej,Lucini:2004my}
that, for the range of $a(\beta)$ involved, these volumes are 
large enough to be, effectively, at zero $T$. For SU(8) 
however, using $8^4$ lattices would not be adequate for the
largest $\beta$-values, as we will see below. (The same is not 
true for the finite $T$ calculation on $8^35$ lattices where 
it is $1/aT$ that sets the scale for finite volume corrections.) 
We therefore take instead the SU(8) calculations on larger 
lattices in
\cite{Lucini:2004my},
and interpolate between the values of $\beta$ used there, to
obtain average plaquettes at the values of $\beta$ we require.
To perform this interpolation we fit with the ansatz
\begin{equation}
\langle u_p \rangle_0(\beta)
=\langle u_p \rangle^{P.T.}_0(\beta)
+\frac{\pi^2}{12}\frac{G_2}{N\sigma^2}(a\sqrt{\sigma})^4
+c_4g^8+c_5g^{10}, \label{eq:interpolate}
\end{equation}
where $\langle u_p \rangle^{P.T.}_0(\beta)$ is the lattice 
perturbative result to ${\cal O}(g^6)$ from 
\cite{Alles:1998is}
and $N=8$.
Our best fit has $\chi^2/{\rm dof}=0.93$ with ${\rm dof}=2$,
and the best fit parameters are $c_4=-6.92$, $c_5=26.15$, and a 
gluon condensate of $\frac{G_2}{N\sigma^2}= 0.72$. 

For the scaling 
of the lattice spacing with $\beta$, needed in \Eq{eq:interpolate} 
and \Eq{eq:final_delta} and in the temperature scale, we used the 
interpolation of $a\sqrt{\sigma}$ as a function of $\beta$, as
given in 
\cite{Lucini:2005vg} 
\footnote{This is excluding the first three $\beta$ values in the 
case of $SU(4)$, which are outside the interpolation regime of 
\cite{Lucini:2005vg}. In that case we have performed a new
interpolation fit to include these points as well. This gave the
string tensions 
$a\sqrt{\sigma}=0.3739(15),0.3440(10),0.3336(10)$ and the derivatives
$-d \log a / d\beta = 1.83(7),1.55(7),1.48(5)$ for
$\beta=10.55,10.60,10.62$.}.
For the temperature scale we need in addition to locate
the value of $\beta$ that corresponds to $T=T_c$ for the
relevant value of $L_t$, and for this we have used the values in 
\cite{Lucini:2003zr,Lucini:2005vg}.
In the case of $SU(3)$ we compared the resulting $T/T_c(\beta)$ with
that of \cite{Boyd:1996bx} where the physical scale was set by
$T_c$. We find that the two functions lie on top of each other for
$L_t=6$. This is consistent with the fact that the $SU(3)$ value of
$T_c/\sqrt{\sigma}$ for $L_t=5,6$ are the same within one sigma
\cite{Lucini:2003zr}. This is true for $SU(8)$ as well, where the
value of $T_c/\sqrt{\sigma}$ for $a=1/(5T_c)$, and $a=1/(8T_c)$, are the
same within one sigma \cite{Lucini:2005vg}, and we find no point to
perform similar comparisons there. For $SU(4)$ the value of
$T_c/\sqrt{\sigma}$ at $a=1/(5T_c),1/(6T_c)$ is $\sim 5$, and $\sim 3.7$
sigma away from the value at $a=1/(8T_c)$ \cite{Lucini:2005vg}, which
may suggest that in this case $T/T_c(\beta)$ at values of $\beta$ that
correspond to $T\simeq 8/5 T_c$ will be smaller when fixing the
physical scale with $T_c$ rather than with the string
tension. Nevertheless the shift between the two is at the level of
$\sim 2\%$, and will not change the results presented here. In
addition, to fix $T/T_c(\beta)$ by fixing $T_c$, requires a larger
scale calculation of $\beta_c(L_t,L_s)$ that will include evaluation
of finite volume corrections, similar to what was done for $L_t=5$ in
\cite{Lucini:2003zr}. In view of the small shifts and the high
calculational price, we shall ignore this potential ambiguity 
in this paper.

\begin{table}[htb]
\caption{Statistics and results of the Monte-Carlo simulations for 
$SU(4)$. \label{table1}}
\begin{ruledtabular}
\begin{tabular}{c|c|c|c|c}
$\beta$ & \multicolumn{2}{l|}{$T>0$} & \multicolumn{2}{l}{$T=0$} \\ \hline 
  & $s_T$ & (lattice sweeps)$\times 10^{-3}$ & $s_0$ & (lattice sweeps)$\times 10^{-3}$ \\ \hline
10.55   &0.537478(84)&      10      &0.537487(81)&      5       \\
10.60   &0.543862(58)&      20      &0.543797(25)&      15      \\
10.62   &0.546212(64)&      10      &0.546068(33)&      10      \\
10.64   &0.550279(70)&      10      &0.548208(16)&      20      \\
10.68   &0.554213(32)&      20      &0.552177(16)&      20      \\
10.72   &0.557649(30)&      20      &0.555861(14)&      20      \\
10.75   &0.560051(27)&      20      &0.558462(13)&      20      \\
10.80   &0.563923(32)&      20      &0.562587(16)&      20      \\
10.85   &0.567592(24)&      20      &0.566453(17)&      20      \\
10.90   &0.571107(17)&      20      &0.570118(16)&      20      \\
11.00   &0.577707(17)&      20      &0.576981(11)&      20      \\
11.02   &0.578985(18)&      20      &0.578279(11)&      20      \\
11.10   &0.583911(20)&      20      &0.583352(12)&      20      \\
11.30   &0.595398(13)&      20      &0.595039(10)&      20      \\

\end{tabular}\end{ruledtabular}
\end{table}

\begin{table}[htb]
\caption{Statistics and results of the Monte-Carlo simulations for 
$SU(8)$. \label{table2}}
\begin{ruledtabular}
\begin{tabular}{c|c|c|c|c|c|c}
 \multicolumn{4}{l|}{$T>0$} & \multicolumn{3}{l}{$T=0$} \\ \hline
 $\beta$ & $s_T$ & $L_s$ & (lattice sweeps)$\times 10^{-3}$ & $\beta$ & $s_0$ & (lattice sweeps)$\times 10^{-3}$ \\ \hline
43.90   &0.525330(80)&   14      &       5       &       43.85   &0.523819(37)&   $>20$     \\
43.93   &0.526873(79)&   8       &       19.5    &       44      &0.528788(18)&   $>20$     \\
44.00   &0.531307(50)&   10      &       $>20$     &       44.35   &0.538491(13)&   $>20$     \\
44.10   &0.534164(34)&   12      &       7       &       44.85   &0.549794(9)&    $>20$    \\
44.20   &0.536650(70)&   14      &       5       &       45.7    &0.565708(4)&    $>20$     \\
44.30   &0.539181(30)&  8      &20     &               &                              &               \\
44.45   &0.542629(38)&   8       &       30      &               &                               &               \\
44.60   &0.545812(35)&   8       &       20      &               &                               &               \\
44.80   &0.549968(37)&   8       &       30      &               &                               &               \\
45.00   &0.553926(38)&   8       &       20      &               &                               &               \\
45.50   &0.562992(28)&   12      &       10      &               &                               &               \\

\end{tabular}\end{ruledtabular}
\end{table}

\begin{table}[htb]
\caption{Statistics and results of the Monte-Carlo simulations for
$SU(3)$. \label{table3}}
\begin{ruledtabular}
\begin{tabular}{c|c|c|c|c}
$\beta$ & \multicolumn{2}{l|}{$T>0$} & \multicolumn{2}{l}{$T=0$} \\ \hline
  & $s_T$ & (lattice sweeps)$\times 10^{-3}$ & $s_0$ & (lattice sweeps)$\times 10^{-3}$ \\ \hline
5.800   &0.568664(100)&      10      &0.567667(29)&      11      \\
5.805   &0.569688(153)&      20      &0.568438(23)&      11      \\
5.810   &0.570624(55)&      10      &0.569218(18)&      11      \\
5.815   &0.571297(81)&      10      &0.569996(26)&      11      \\
5.820   &0.572205(78)&      10      &0.570788(16)&      11      \\
5.900   &0.583058(38)&      10      &0.581854(20)&      11      \\
6.150   &0.609377(27)&      10      &0.608971(8)&      11      \\
6.200   &0.613966(31)&      10      &0.613628(13)&      11      \\
\end{tabular}\end{ruledtabular}
\end{table}

\subsection{Finite volume effects}

For $N=4,8$, one is able to use lattice volumes much
smaller than what one needs for $SU(3)$ 
\cite{Boyd:1996bx}. 
That this is so for the deconfinement transition, has been 
explicitly demonstrated in
\cite{Lucini:2003zr,Lucini:2005vg},
and is theoretically expected, much more generally, as $N\to\infty$.
The main remaining concern has to do with tunnelling between confined
and deconfined phases near $T_c$. When $V\to\infty$ tunnelling
occurs only at $\beta=\beta_c$ (in a calculation of sufficient 
statistics) and the system is in the appropriate 
pure phase for $T<T_c$ and for $T>T_c$. On a finite volume, where 
this is no longer true, one minimises finite-$V$ corrections
by calculating the average plaquettes only in field configurations
that are confining, for $T<T_c$, or deconfining, for $T>T_c$.
This ensures that the system is as close as possible to being
`large and homogeneous' as is required in the derivations
of this Section. Because the latent heat grows $\propto N^2$
\cite{Lucini:2005vg}
the region $\delta T$ around $T_c$ in which there is significant
tunnelling shrinks as $\delta T\propto 1/N^2$ for a given $V$.
Hence we can reduce $V$ as $N$ increases without increasing the
ambiguity of the calculation. For SU(3), where the phase transition 
is only weakly first order, frequent tunneling occurs 
in the vicinity of $T_c$ in the volume we use, and it is not
practical to attempt to separate phases. This will smear the apparent variation of the pressure across $T_c$ in the
case of SU(3).

We now turn to a more detailed discussion of finite volume effects.
If $\xi$ is the longest correlation length, in lattice units, in 
a volume of length $L$, then finite volume effects will be
negligible if $\xi \ll L$. In addition finite volume corrections
will be suppressed as $N\to\infty$. In our particular context, 
$\xi$ is given by the inverse mass of the lightest (non-vacuum) 
state that couples to the loop that winds around the temporal torus.
In both the confined and deconfined phases, these masses 
decrease as $T\to T_c$. Therefore the largest length scale is set 
by the masses at $T=T_c$. As $N$ increases these masses increase
towards their limits, with $1/N^2$ corrections that are quite
large
\cite{Lucini:2005vg}.

\subsubsection{The deconfined phase}

In the deconfined phase, on an $L_s^3\times 5$ lattice at $T=T^+_c$, 
the value of $\xi$ is about $12.5$ lattice spacings for $SU(3)$, 
while it is about $5.2$, and $2.4$ lattice spacings for $SU(4)$, 
and $SU(8)$ respectively \cite{Lucini:2005vg}. This suggests that our choice of $L_s=16$ 
for SU(4) and $L_s=8$ for SU(8) should be adequate. In addition it 
is known from calculations of $T_c$
\cite{Lucini:2002ku,Lucini:2003zr,Lucini:2005vg}
that on such lattices the tunnelling is sufficiently rare that 
even at $T=T_c$ one can reliably categorise field configurations 
as confined or deconfined and hence calculate the average plaquette 
in just the deconfined phase if one so wishes. For our supplementary
SU(3) calculations we use $L_s=20$ which is much smaller in units of 
$\xi$. In practice this means that in this case we are unable to
separate phases at $T\simeq T_c$.

To explicitly confirm our control of finite volume effects,
we have compared the SU(8) value of $\langle u_p(\beta) \rangle$ as
measured in the deconfined phase of the our $8^3\times 5$ lattice 
with other $L_s^3\times 5$ results from other studies 
\cite{Bringoltz:2005xx}.
As summarised in Table~\ref{table4}, the results are consistent
at the $2\sigma$ level.

\begin{table}[htb]
\caption{Finite volume effects for plaquette average in the deconfined 
phase on a $L_t=5$ lattice, for $SU(8)$. 
\label{table4}}
\begin{ruledtabular}
\begin{tabular}{c|c|c|c|c}
$\beta$ & $L_s=8$ & $L_s=10$ & $L_s=12$ & $L_s=14$ \\ \hline
43.95   &       -       &0.529788(100)&  0.529944(65)&   -               \\
44.00   & 0.531343(45)    &0.531307(50)&   -       &       -       \\
44.10   &0.534219(54)    &       -       &0.534164(34)&   -       \\
44.20   &0.536714(33)    &       -       &0.536689(54)&   0.536650(70)\\
44.25   &       -       &       -       &0.537954(60)&   0.537850(100)\\
44.30   &0.539181(29)    &       -       &       -       &0.539220(100)\\
45.50   &0.563093(41)    &       -       &0.562992(28)&   -       \\
\end{tabular}\end{ruledtabular}
\end{table}

\subsubsection{The confined phase}

As we remarked above (see below for explicit evidence) we have 
$\langle u_p \rangle_T \simeq \langle u_p \rangle_0$  in the
confined phase and so  the contribution in \Eq{eq:pint2} of the range 
of $\beta$ where the finite $T$ system is confined is very small.
Nonetheless, we include an integration over that range for
completeness and so we need to discuss possible finite $V$ corrections 
for this case as well.

In the confined phase, on an $L_s^3\times 5$ lattice at $T=T^{-}_c$, 
the value of $\xi$ is about $9.5$ lattice spacings for $SU(3)$, but 
drops to about $5$ and $3.5$ for $SU(4)$ and $SU(8)$ respectively
\cite{Lucini:2005vg}.
This leaves our choice  of $L_s$ still reasonable for $SU(4)$ but
somewhat worse for SU(8). In Table~\ref{table5} we provide a 
finite volume check for the latter case that proves reassuring.

Finally we return to our earlier comment that for the `T=0'
$L^4$ lattice calculations, a size $L=8$ in SU(8) would not be 
large enough. This is demonstrated, for our largest $\beta$-value,
in Table~\ref{table6}, where we also present the value of $L_t\times T/T_c(\beta)$ (in our $L_t=5$ calculations). In the confined $L_s^4$ lattice, finite volume effects will be suppressed when the latter is much smaller than $L_s$. Clearly for $\beta=45.70$, and $L_s=8$, this is not the case.

By contrast, for $SU(4)$ the finite volume effects seems not to be
large on the $16^4$ lattice as we checked for our largest value of
$\beta=11.30$. There the value of the plaquette on a $20^4$ lattice is
$0.595014(4)$ \cite{Lucini:2001ej}, which is consistent within $\sim
2.3$ sigma with the value presented in Table~\ref{table1}. This is in
spite of the fact that for this coupling $L_t\times T/T_c=10$, and is not so much smaller than $L_s=16$.    

\begin{table}[htb]
\caption{Finite volume effects for plaquette average in the confined 
phase on a $L_t=5$ lattice, for $SU(8)$.
\label{table5}}
\begin{ruledtabular}
\begin{tabular}{c|c|c|c|c}
$\beta$ & $L_s=8$ & $L_s=10$ & $L_s=12$ & $L_s=14$ \\ \hline
43.90   &      0.525750(87)      &       -        &       0.525613(54)&      0.525425(90)\\
43.95   &       -                       &       0.527240(34)&      0.527275(48)&      0.527280(50)\\
44.00   &       -                       &       -                       &       0.528867(33)&      0.528810(50)\\
44.10   &       -                       &       -                       &       0.531880(45)&      0.531900(60)\\
\end{tabular}\end{ruledtabular}
\end{table}

\begin{table}[htb]
\caption{Finite volume effects for plaquette average in the confined 
phase on a $L^4$ lattice, for $SU(8)$. The last column is for $L_t=5$.
\label{table6}}
\begin{ruledtabular}
\begin{tabular}{c|c|c|c|c}
$\beta$ & $L_s=8$ & $L_s=10$ & $L_s=16$ & $L_t\times T/T_c$ \\ \hline
44.00   &       0.528876(39)&      0.528788(18)&      -                       &       5.05    \\
45.70   &       0.566089(23)&      -                       &       0.565708(4)&      8.20    \\
\end{tabular}\end{ruledtabular}
\end{table}

\section{Results}
\label{results}

To obtain the pressure from the values of the average plaquette
presented in Tables~\ref{table1}--\ref{table3}
we need to perform the integration in \Eq{eq:pint2}, which we 
do by numerical trapezoids. We have already remarked 
that the contribution to the pressure from the confined
phase is negligible. In Table~\ref{table7} we provide some accurate evidence
for this. We show the values of the average plaquette on
$L^4$ lattices, corresponding to $T\simeq 0$, as well as the
values on $L_s^3 5$ lattices at $T\simeq T_c$, with the latter 
obtained separately in the confined and deconfined phases. 
(These volumes are large enough for there to be no tunnelling,
or even attempted tunnelling, within our available statistics.)
We see that for both $SU(4)$ and $SU(8)$ there is no visible
difference between the plaquette at $T=0$ and $T=T_c$
in the confined phase at, say, the $2\sigma$ level. Any difference,
(and there obviously must be some difference)
is clearly negligible when compared to the difference between
the confined and deconfined phases at (and above) $T_c$.

\begin{table}[htb]
\caption{The plaquette average in the confined phase, $C$, 
at $T\simeq T_c$ compared to the $T=0$ value and to the
value in the deconfined phase, $D$. For $SU(4)$ and  $SU(8)$. 
\label{table7}}
\begin{ruledtabular}
\begin{tabular}{c|c|c|c|c|c}
$\beta$ & $N$ & $lattice$ & $\langle u_p \rangle$ & $phase$ & $T$\\ \hline
10.635  &  4  &  $32^3 5$ & 0.549563(33) &  D &  $T_c$   \\
        &     &  $32^3 5$ & 0.547689(11) &  C &  $T_c$   \\
        &     &  $10^4$   & 0.547640(27) &  C &  $0$   \\ \hline
43.965  &  8  &  $12^3 5$ & 0.530352(23) &  D &  $T_c$   \\
        &     &  $12^3 5$ & 0.527725(27) &  C &  $T_c$   \\
        &     &  $10^4$   & 0.527648(24) &  C &  $0$   \\ \hline
\end{tabular}\end{ruledtabular}
\end{table}

In presenting our results for the pressure, we shall 
normalize to the lattice Stephan-Boltzmann result given by
\begin{equation}
\left( p/T^4
\right)_{\text{free--gas}}=(N_c^2-1)\frac{\pi^2}{45} \times R_I(L_t). \label{eq:free}
\end{equation}
Here $R_I$ includes the effects of discretization errors in the integral method \cite{Engels:1999tk,Juergen}. For large values of $L_t$, and an infinite volume, it is given by
\begin{equation}
R_I(L_t)=1+
\frac{8}{21}\left(\frac{\pi}{L_t}\right)^2+
\frac{5}{21}\left(\frac{\pi}{L_t}\right)^4+{\cal O}
\left(\frac1{L_t}\right)^6. \label{eq:R_I}
\end{equation}

Since some values of $L_t$ discussed in this work are not very large, we shall use the full correction, which includes higher orders in $1/L_t$, instead of \Eq{eq:R_I}. This was calculated numerically for the infinite volume limit in \cite{Engels:1999tk} for $L_t=4,6,8$, and we supplement this calculation, with the same numerical routines \cite{Juergen}, for other values of $L_t$. A summary of $R_I(L_t)$ in the infinite volume limit is given in Table~\ref{table8}. 

\begin{table}[htb]
\caption{The lattice discretisation errors correction factor $R_I(L_t)$ in the infinite volume limit.
\label{table8}}
\begin{ruledtabular}
\begin{tabular}{c|c|c|c|c|c}
$L_t=2$ & $L_t=3$ & $L_t=4$ & $L_t=5$ & $L_t=6$ & $L_t=8$\\ \hline
$2.04526(4)$ & $1.6913(2)$ & $1.3778(1)$  & $1.2129(6)$ & $1.1323(1)$ & $1.0659(1)$
\end{tabular}\end{ruledtabular}
\end{table}

We find that the full correction for $L_t=5$ is a $\sim 21\%$ effect, which, without
this normalisation, might obscure the physical effects that
we are interested in. 
This is an appropriate normalisation
because we expect Eq.~(\ref{eq:free}) to provide the $T\to\infty$
limit of $p/T^4$. The same applies to the internal energy 
density, since $\epsilon \to 3p$ as  $T\to\infty$, and
so when presenting our results for  $\epsilon/T^4$ we normalise
it with the expression in Eq.~(\ref{eq:free}). For similar
reasons we shall use the same normalisation when presenting
our results for the entropy. For $\Delta/T^4$ it is less clear
what normalisation one should use since  
$\Delta = \epsilon -3p \to 0$ as $T\to \infty$, but for ease 
of comparison we shall once again normalise using Eq.~(\ref{eq:free}).

To facilitate the comparison of our results with 
earlier work on SU(3)
\cite{Boyd:1996bx}, 
which was done for $L_t=4,6,8$, we have performed  $SU(3)$
simulations with $L_t=5$. The spatial size is $L_s=20$ 
which should be  sufficiently large in the light of our 
above discussion of finite volume effects (and we note that 
it satisfies an empirical rule that one needs $L_s/L_t\ge 4$ 
\cite{Engels:1990vr}).

\begin{figure}[htb]
\includegraphics[width=12cm]{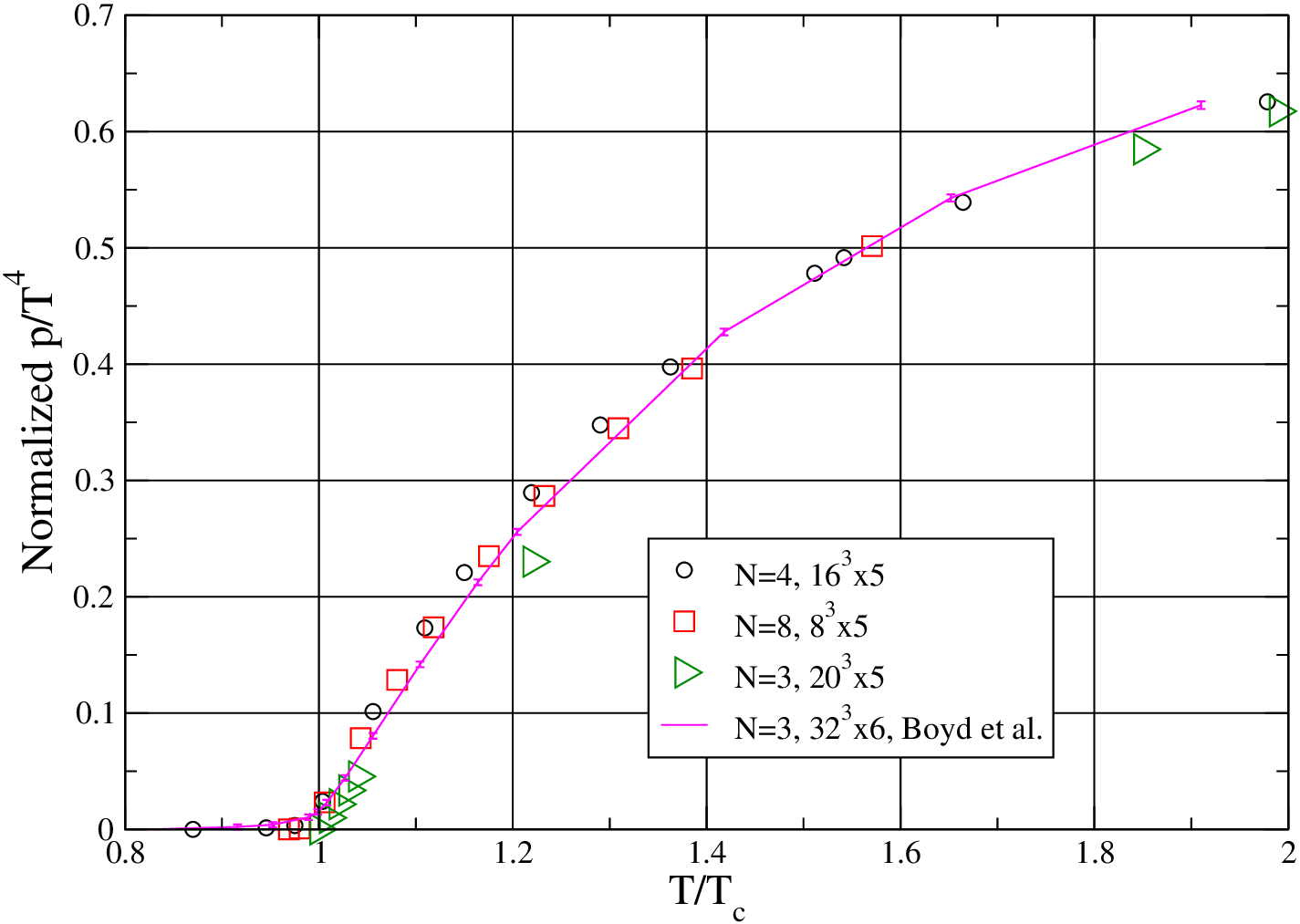}
\vspace{0.5cm}
\caption{The pressure, normalized to the lattice Stephan-Boltzmann 
pressure, including the full discretization errors. The symbol's vertical sizes are representing the largest error bars (which are received for the highest temperature). The 
solid line is for $SU(3)$ and $L_t=6$ from 
\cite{Boyd:1996bx}.} \label{fig1}
\end{figure}

We present our $N=4$ and  $N=8$ results for $p/T^4$ in 
Fig.~\ref{fig1}. We also show there our calculations of the $SU(3)$
pressure for $L_t=5$, as well as the $L_t=6$ calculations from 
\cite{Boyd:1996bx}.
Although our errors on the $SU(3)$ pressure are probably 
underestimated, since the mesh in $\beta$ is quite coarse,
nonetheless one can clearly infer that the pressure 
in the $SU(4)$ and $SU(8)$ cases is remarkably close to that
in $SU(3)$ and hence that the well-known pressure deficit 
observed in SU(3) is in fact a property of the large-$N$ 
planar theory.

\begin{figure}[htb]
\includegraphics[width=12cm]{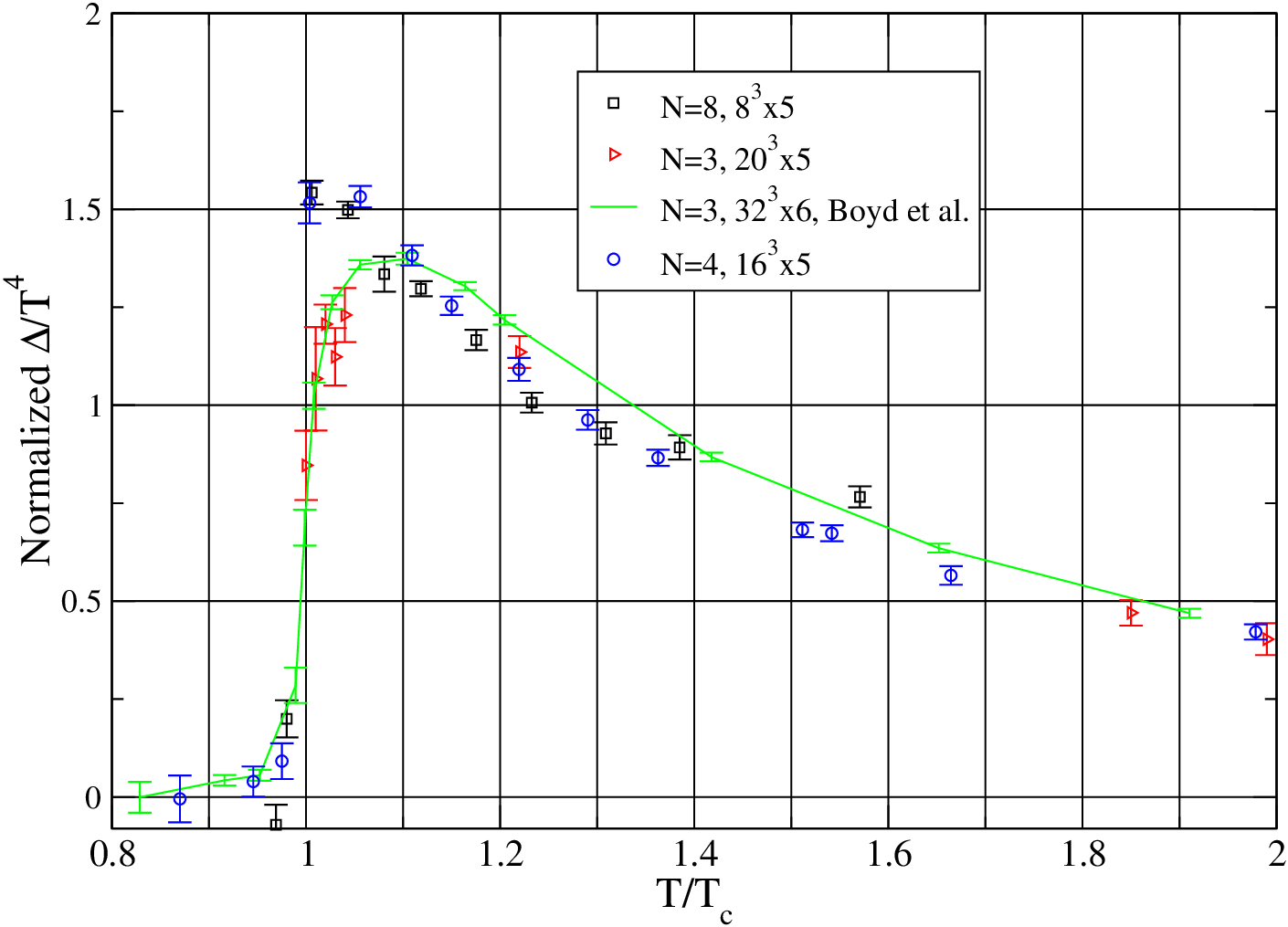}
\vspace{0.5cm}
\caption{Results for $\Delta(T)/T^4=T\frac{\partial p/T^4}{\partial T}$,
 normalized by the same coefficient as we normalize the pressure. The solid line is
for $SU(3)$ and $L_t=6$ from \cite{Boyd:1996bx}.} \label{fig2}
\end{figure}

In Fig.~\ref{fig2} we present our results for  $\Delta/T^4$ as 
calculated from \Eq{eq:final_delta}.
This quantity can be considered as a measure of the interaction and 
non-conformality of the theory, since it 
is identically zero both for the noninteracting Stephan-Boltzmann
case, and for the ${\cal N}=4$ supersymmetric $SU(N)$ gauge theory. 
As remarked above, we normalise with the expression in
Eq.~(\ref{eq:free}). We also note that in this case there are no
errors from a  numerical integration, and this enables a fair
comparison with the $SU(3)$ data of 
\cite{Boyd:1996bx}.
Comparing the results for different $N$ we
see that, just as for the pressure, the results for all these 
gauge theories are very similar.

To see what is the behaviour of $\Delta/T^4$ at even higher
temperatures, we use the plaquette averages on lattices with 
$L_t=2,3,4,5$, that have been calculated at fixed couplings 
which correspond to $T\simeq T_c$ for  $L_t=5$
\cite{Lucini:2005vg}. 
We present the results in Table~\ref{table9}. For the evaluation 
of $\Delta$ one needs $d\log a/ d\beta$ which we present in the 
table as well.

\begin{table}[htb]
\caption{Plaquette average in the deconfined phase for lattice with 
fixed coupling, at different values of $L_t$, and with $\beta$ that 
corresponds to roughly the deconfining temperature at $L_t=5$: 
$\beta=5.800,10.635,44.00$ for $N=3,4,8$. The data for $L_t=5$ are 
obtained for $L=64,32,10$ for $N=3,4,8$ (for $N=3$, $\delta \<u_p \> $ 
is the difference between the plaquette as calculated within 
separate  confined and deconfined sequences of field configurations).
\label{table9}}
\begin{ruledtabular}
\begin{tabular}{c|c|c|c|c|c|c}
$N$ & $L^3\times 5$ & $8^3\times 4$ & $8^3\times 3$ & $8^3\times 2$ & $10^4$ & $-d\log a/d\beta$\\ \hline
 3 &   $\delta \< u_p \> =0.00080(5)$  &0.570987(37)&  0.573311(34)&   0.578121(27) & 0.567642(29) & 2.075(17)           \\ \hline 
4  & 0.549563(33)    & 0.551604(33) &   0.554047(27)  &  0.559163(24)  & 0.547640(27) & 1.440(23)   \\ \hline
8  & 0.531202(92)    & 0.533066(25) &   0.535991(24) &  0.541518(17)  &  0.528788(18) & 0.384(20)     \\
\end{tabular}\end{ruledtabular}
\end{table}

In such calculations where one varies $T$ by varying $L_t$, 
the lattice spacing varies as $a = 1/L_t \times 1/T$ when 
expressed in units of the relevant temperature scale, and so 
lattice spacing corrections will vary with $T$. 

\begin{figure}[htb]
\includegraphics[width=12cm]{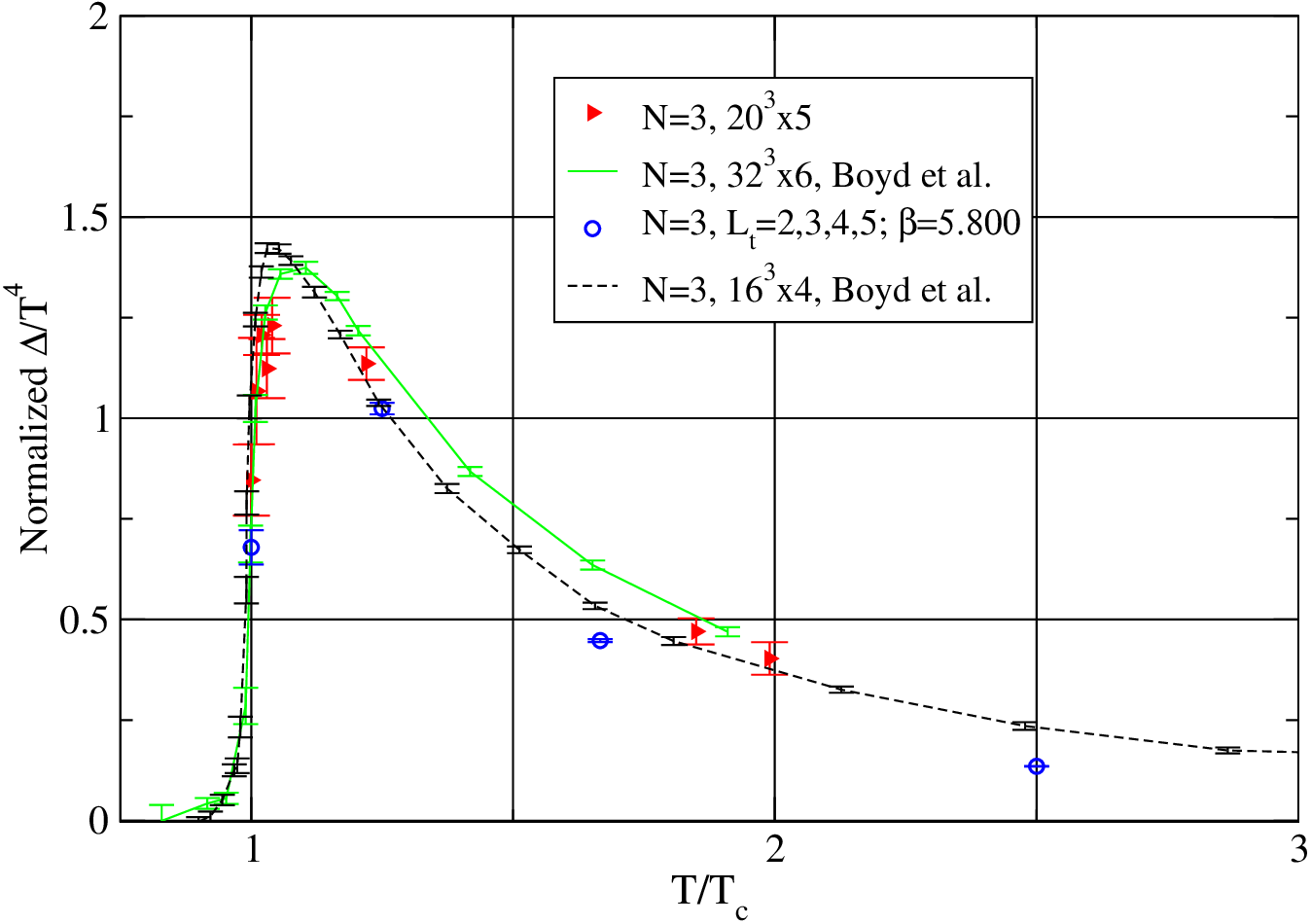}
\vspace{1cm}
\caption{Results for $\Delta(T)/T^4=T\frac{\partial p/T^4}{\partial T}$,
 normalized to the free-gas result. The lines are
for $SU(3)$ and $L_t=4,6$ from \cite{Boyd:1996bx}. Red triangles correspond to $L_t=5$, and changing $\beta$, while blue circles correspond to changing $L_t$ and keeping a fixed $\beta=5.800$.} \label{fig3}
\end{figure}

The resulting values of $\Delta$ in the case of SU(3) are 
plotted in Fig.~\ref{fig3} where they are compared
to the results obtained from calculations where 
one varies $T$ by varying $\beta$ at fixed  $L_t$. These
calculations include ours for $L_t=5$ and those of
\cite{Boyd:1996bx}
for $L_t=4,6$. 

As we see from Fig.~\ref{fig3} our $L_t=5$  $SU(3)$  results do 
in fact lie between the $L_t=4,6$ results of \cite{Boyd:1996bx}
as one would expect. 
We observe that the $T$ dependence is very
similar in all cases, and that the remaining $L_t$
dependence appears to be much the same for the different kinds of
calculation. This gives us confidence that performing calculations
where we vary $T$ by varying $L_t$ at fixed $\beta$ does not
introduce any unanticipated and important systematic errors.

\begin{figure}[htb]
\vspace{1.0cm}
\includegraphics[width=12cm]{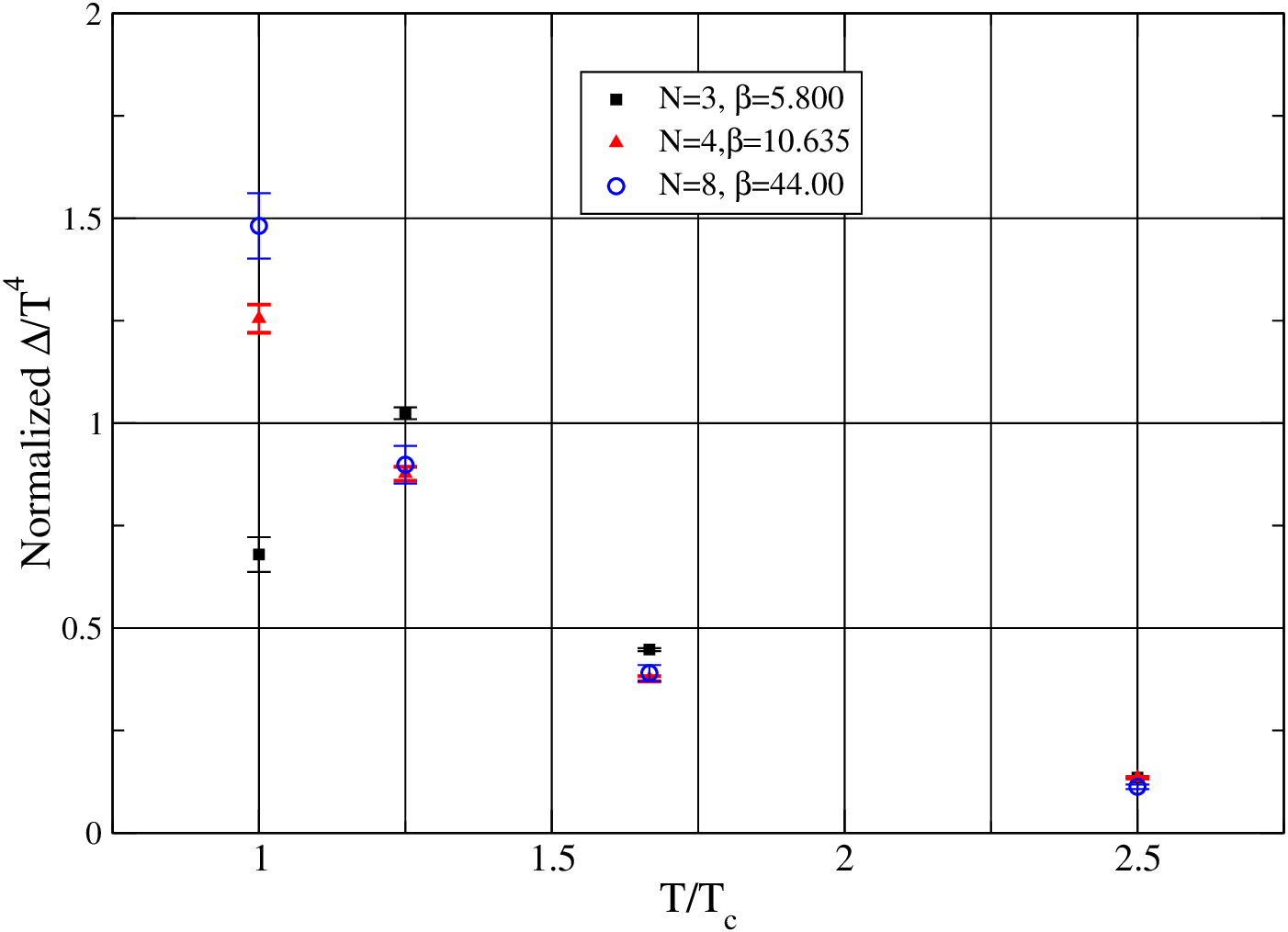}
\vspace{0.5cm}
\caption{Results for $\Delta(T)/T^4=T\frac{\partial p/T^4}{\partial T}$ for $N=3,4,8$, by fixing $\beta=\beta_c(L_t=5)$, while changing $L_t=2,3,4,5$.} \label{fig4}
\end{figure}

Having performed this check, we compare 
in Fig.~\ref{fig4} our results for $\Delta$ in the
range  $T_c \leq T \leq 2.5T_c$ that corresponds to  
$5 \geq L_t \geq 2$. This comparison confirms what we observed in
Fig.~\ref{fig2} over a smaller range of $T$: $\Delta$ is
very similar for all the values of $N$ (except very close to $T_c$), implying that this
is also a property of the $N=\infty$ planar limit.

\begin{figure}[htb]
\includegraphics[width=12cm]{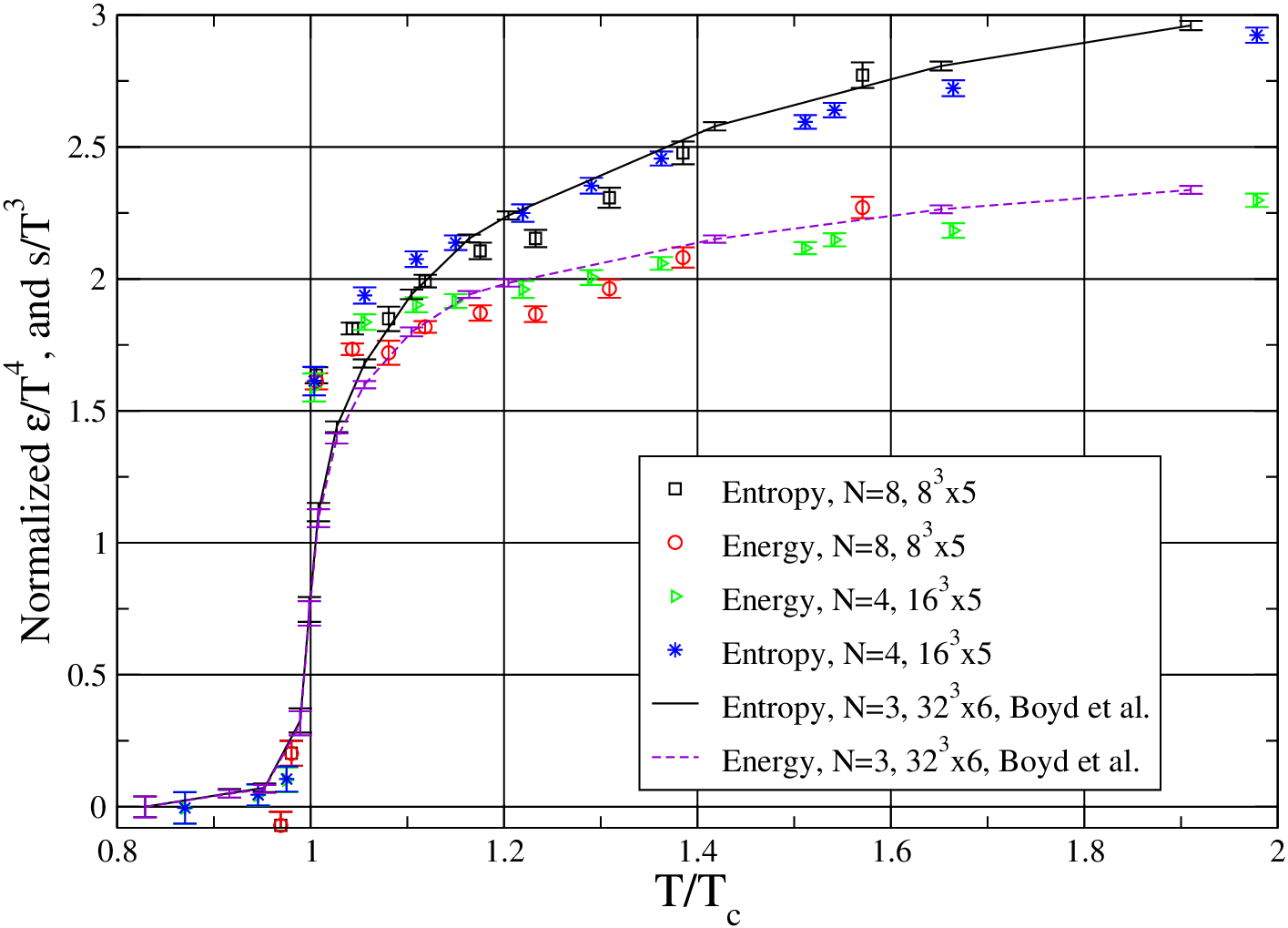}
\vspace{0.5cm}
\caption{Results for energy density and entropy, normalized to the 
lattice Stephan-Boltzmann result, including the full discretization errors. The solid line is for $SU(3)$ 
and $L_t=6$ from \cite{Boyd:1996bx}.} \label{fig5}
\end{figure}

\vspace{0.5cm}
Finally we present in Fig.~\ref{fig5} our results for the normalized 
energy density $\epsilon=\Delta+3p$, and the entropy per unit volume 
$s=(\epsilon+p)/T$. The lines are the $SU(3)$ result of 
\cite{Boyd:1996bx} 
with $L_t=6$. Again we see very little dependence on the gauge group,
implying very similar curves for $N=\infty$.

\section{Summary and discussion}
\label{summary}

In this work we have analyzed numerically the bulk thermodynamics of 
SU(4) and SU(8) gauge theories. We found that the pressure, when 
normalized to the Stephan-Boltzmann lattice pressure, is practically 
the same as for SU(3), in the range $T_c\leq T \leq 1.6T_c$ that we 
analyze. We found the same to be the case for the internal energy
and entropy, as well as for the quantity $\Delta = \epsilon - 3p$
(where we were able to explore temperatures up to $T\simeq 2.5T_c$).
All this implies that the dynamics that drives the deconfined 
system far from its noninteracting gluon plasma limit, must remain
equally important in the $N=\infty$ planar theory. This is encouraging 
since that limit is simpler to approach analytically, in
particular using gravity duals. 

Our results have been (mostly) obtained for lattice spacings $a=1/(5T)$ 
and it would be useful to perform a larger scale calculation that
allows us to perform an explicit continuum extrapolation. 
However past SU(3) calculations of the pressure, and calculations
in SU($N$) of various physical quantities, strongly suggest
that our choice of $a$ already provides us with a reliable
preview of what such a more complete calculation would produce.

Our results imply that any explanation of the QCD pressure deficit 
must survive the large--$N$ limit, and so should not be driven by 
special features particular to SU(3). This can provide a strong 
constraint on such explanations. For example, in approaches based 
on higher order perturbation theory, it tells us that the important 
contributions must be planar. In models focussing on resonances and
bound states, it must be that the dominant states are coloured,
since the contribution of colour singlets will vanish as $N\to\infty$.
Models using `quasi-particles' should  place these in
colour representations that do not exclude their presence at 
$N=\infty$, and in fact give them $T$-dependent properties which depend weakly on $N$. Also, topological fluctuations  should play no role in this   
deficit since the evidence is that there are no topological 
fluctuations of any size in the deconfined phase at large-$N$
\cite{Lucini:2004yh,DelDebbio:2004rw}. 

Finally, we emphasize that our conclusion that the SU(3) pressure 
and entropy deficits are features of the large-$N$ gauge theory, 
means that these `observable' phenomena can, in principle, be 
addressed using AdS/CFT  gravity duals. Indeed it is
precisely where the deficit is large that the
coupling must be strong and this is also precisely 
where, at large $N$, such dualities can be established.
As has been frequently emphasized (see for example 
\cite{Gavai:2004se,Gavai:2005da}) 
the deficit in the normalized entropy is not far from the value 
of $s/s_{\text{free-gas}}=3/4$ given by the AdS/CFT prediction. 
In this paper we have found that large-$N$ gauge theories show the same 
behaviour, as we see in Fig.~\ref{fig5} where, for the entropy, the 
horizontal line $s_{\text{normalized}}/T^3=3$ would correspond 
to $s/s_{\text{free-gas}}=3/4$. Our results can therefore
serve as a bridge between the AdS/CFT approach to large-$N$ and
the observable world of QCD.

\begin{acknowledgments}

We are thankful to Juergen Engels for useful discussions on the finite lattice spacing corrections of the free gas pressure in the integral method, and in particular for giving us the numerical routines to calculate them. 
Our lattice calculations were carried out on PPARC 
and EPSRC funded computers in Oxford Theoretical 
Physics. BB acknowledges the support of a PPARC
postdoctoral research fellowship.

\end{acknowledgments}

\bibliography{pressure}

\end{document}